\documentclass{appolb}
\usepackage{graphicx}
\begin{document}
\title{Probing the QCD phase diagram with HBT femtoscopy%
\thanks{Presented at Epiphany conference 2022, Krakow}%
}
\author{S\'andor L{\"o}k{\"o}s
\address{Institute of Nuclear Physics, PAS \\
Hungarian University of Agriculture and Life Sciences \\ 
E{\"o}tv{\"o}s Lor\'and University}
\\[3mm]
%{Third Author % of different affiliation
%\address{affiliation}
%}
%\\[3mm]
%the Name(s) of other Author(s)
%\address{affiliation}
}
\maketitle
\begin{abstract}
Intensity interferometry originates from the field of radio astronomy on the trace of Robert Hanbury Brown and Richard Quincy Twiss. In high energy physics, the phenomena was discovered by Goldhaber, Goldhaber, Lee and Pais. In radio astronomy the method is exceeded by more modern approaches but in high energy physics, the measurements of Hanbury Brown-Twiss (HBT) or Goldhaber-Goldhaber-Lee-Pais (GGLP) type of correlations are important tools to access the spatio-temporal properties of the matter under extreme conditions on subatomic scales. In this paper, I review recent experimental results from energies of the Relativistic Heavy Ion Collider (RHIC) to the Large Hadron Collider (LHC) and discuss their possible implication on the equation of state of the QCD matter.
\end{abstract}
  
\section{Introduction}
Intensity interferometric phenomena was observed by Robert Hanbury Brown and Richard Quincy Twiss reported for the first time in Ref. \cite{HanburyBrown:1956bqd}. In proton--antiproton annihilation, intensity correlation among identically charged pions was observed by Goldhaber, Goldhaber, Lee and Pais  while they searched for the $\rho$ meson. They could explain their experimental results on the basis of the Bose-Einstein symmetrization of the wave-function of the like-charged pions.

The HBT or Bose-Einstein correlation functions are related to the Fourier-transform of the source distribution, therefore these type of measurements can be used to access the spatio-temporal properties of the matter and reveal underlying processes on the femtometer scale. Hence, the field of the Bose-Einstein correlations often referred as femtoscopy. The term was coined by Richard Lednick\'y in Ref. \cite{Lednicky:2001qv}.

The detailed shape of the Bose-Einstein correlation functions with recent experimental resolutions can be investigated with unprecedented precision. New results from RHIC to LHC energies have showed that the usual Gaussian or Cauchy assumptions cannot characterize the measured experimental data in a statistically acceptable way.

In this paper, I introduce the Bose-Einstein correlation functions, the motivation behind the L\'evy parametrization and the possible physical interpretation of a new parameter. I also give an overview of recent experimental results which support the idea of the L\'evy parametrization from RHIC to LHC energies.

\section{Bose-Einstein correlation function}

The general definition of the $2$-particle correlation function can be written in terms of invariant momentum distributions as
\begin{equation}
    C_2(p_1,p_2)=\frac{N_2(p_1,p_2)}{N_1(p_1)N_1(p_2)},
\end{equation}
where $N_i$ are the $i$-particle momentum distribution functions. I will discuss 2-particle correlations but let me also emphasis that there are 3-particle HBT measurements as well (e.g. in Ref. \cite{Kurgyis:2019xzt}). The definition above can be generalized to $n$-particle correlations straightforwardly.

The momentum distributions can be related to the particle emitting source via the Yano-Koonin formula \cite{Yano:1978gk}
\begin{equation}
    N_1=\int S(x,p)\Psi(x,p)
\end{equation}
\begin{equation}
    N_2(p_1,p_2)=\int d^4 x_1, d^4 x_2 S(x_1,p_1)S(x_2,p_2) |\Psi^{(2)}_{p_1,p_2}(x_1,x_2)|^2,
\end{equation}
where $S$ denotes the source functions and $\Psi^{(2)}$ is the pair wavefunction. Since the two-particle Bose-Einstein correlation function is related to the pair distribution, the 2-particle source function can be introduced via the single particle source functions as
\begin{equation}
    D(x,K)=\int d^4 r S(r+x/2,K) S(r-x/2,K),
\end{equation}
where $K=0.5(p_1+p_2)$ the average momentum of the pair and $r$ is the four-vector of the center of mass of the pair. If we assume the wave-function to be plane wave then the two particle correlation function can be written in the following form:
\begin{equation}
    C_2(Q,K)=1+\frac{\tilde{D}(Q,K)}{\tilde{D}(0,K)}.
    \label{eq:C20}
\end{equation}
Here, $Q$ is the relative momentum variable in one-dimension defined in the longitudinal comoving system (LCMS) as
\begin{equation}
    Q = \sqrt{\left(p_\textmd{1,x}-p_\textmd{2,x}\right)^2 + (p_\textmd{1,y}-p_\textmd{2,y})^2+q^2_\textmd{long,LCMS}}
\end{equation}
where
\begin{equation}
    q^2_\textmd{long,LCMS}=\frac{4(p_\textmd{1,z}E_2-p_\textmd{2,z}E_1)^2}{(E_1+E_2)^2-(p_{1,z}+p_{2,z})^2}.
\end{equation}

The definition of the Bose-Einstein correlation function in Eq. (\ref{eq:C20}) assumes that there is a single source emitting particles incoherently. This picture would imply that the correlation function should be equal to 2 at zero relative momentum, if final state interactions are negligible. In practice, the precision of the momentum resolution only allows to measure $Q=4-8$ MeV/c relative momenta and the $Q=0$ can be reached via extrapolation.

In the core-halo picture, the particle emitting source is assumed to be a composite one: a core part which emits the primordial particles, and a halo part which consists of particles coming from long-lived resonance decays. The details of the model can be found in Ref. \cite{Csorgo:1994in}. Here, we only mention that if one take into account the contribution of the halo part, it introduces the $\lambda$ intercept parameter into the definition of the correlation function as
\begin{equation}
    C_2(Q,K)=1+\lambda(K)\frac{\tilde{D}_{c,c}(Q,K)}{\tilde{D}_{c,c}(0,K)}
\end{equation}
where the $_{c,c}$ indexes denote that only the pure core parts are considered of the sources. For the details, see Ref.  \cite{Csorgo:1994in}.

Experimentally, charged particles are measured, so one should take into account for their final state interactions, such as the Coulomb repulsion of the like-charged particles. The widely used method to consider the Coulomb effect is the Bowler-Sinyukov method \cite{1991PhLB27069B,Akkelin:1995gh} which has the form of
\begin{equation}
    C_2(q,K)=1-\lambda+\lambda \int d^3\textbf{r} D_{c,c} (\textbf{r},K)|\psi^{(2)}_{\textbf{q}}(x_1,x_2)|^2,
\label{eq:generalC2}
\end{equation}
where the two-particle Coulomb-interacting wave-function, which is not a plane wave anymore, with the Coulomb-parameter $\eta_c=\frac{m_\pi c^2 \alpha_{f.s}}{2\hbar q c}$, is
\begin{equation}
    \Psi^{(2)}_{\textbf{q}}(x_1,x_2)=\frac{\Gamma(1+i\eta_c)}{\sqrt{2} e^{\pi\eta_c/2}} \left[ e^{i\textbf{qr}}F(-i\eta_c,1,i(kr-\textbf{qr}))+(\textbf{r} \leftrightarrow -\textbf{r}) \right].
\end{equation}
Here, $F(\cdot,\cdot,\cdot)$ is the confluent hypergeometric function, $\Gamma(\cdot)$ is the Gamma-function and $\alpha_{f.s}$ is the fine structure constant.

For point-like sources which can describe by the Dirac $\delta$-function, the Coulomb factor would be simply the Gamow factor. For more complex sources such as the L\'evy distribution, numerical methods should be applied as it is detailed in Refs. \cite{PHENIX:2017ino,Csanad:2019lkp}.

The role of the strong final state interaction in presence of Lévy sources has been investigated in Ref. \cite{Kincses:2019rug}. An Aharonov-Bohm like effect also could be considered which could cause a specific transverse mass dependence of the intercept parameter. For details, see Ref. \cite{Csanad:2020qtu}.

\section{L\'evy parametrization}

Experimental results indicate that the shape of the Bose-Einstein correlation functions contain a power-law-like long-range component which cannot be characterized by the Gaussian distribution, instead L\'evy-type of distribution should be utilized. Recently it has been shown too that L\'evy-shaped sources appear in phenomenological, hydrodynamics-based models on an event-by-event basis as well \cite{Kincses:2022eqq}. The one-dimensional, symmetric version of such distribution is defined with a Fourier-transform as
\begin{equation}
    \mathcal{L}(\alpha,R,\textbf{r}) = \int d^3\textbf{q}e^{i\textbf{qr}}e^{-\frac{1}{2}|R\textbf{q}|^\alpha},
\end{equation}
where $R$ is the L\'evy scale parameter which is not equivalent with the Gaussian width and $\alpha$ is the so-called L\'evy index of stability or L\'evy exponent. It can be seen from this general form if $\alpha=1$ the Cauchy distribution can be restored that and the $\alpha=2$ corresponds to the Gaussian case.

Besides several reason (QCD jets, anomalous diffusion), the L\'evy assumption motivated by the possible relation of the L\'evy exponent and the critical exponent $\eta$ of a system at a second order phase transition, as it is suggested in Ref. \cite{Csorgo:2005it}. The $\eta$ exponent characterizes the power-law behavior of the spatial correlation at the critical point. An order parameter's correlation function (in three dimensions, as a function of distance $r$) will have the exact same form as the limiting behavior of the L\'evy source distribution.

From statistical physics it is also known, that the universality class of the second order QCD phase transition is the same as the 3D Ising model's (see in Ref. \cite{Stephanov:1998dy}). If the above mentioned relation is valid between $\eta$ and $\alpha$ then in vicinity of the hypothetical critical point in the QCD matter sudden changes should rise for the value of $\alpha_\textmd{L\'evy}$ or specially $\alpha_\textmd{L\'evy} = \eta_\textmd{Ising} = 0.50 \pm 0.05$ could be observed (see Ref. \cite{PhysRevB.52.6659}). This possible relation between the shape of the Bose-Einstein correlation functions and the critical behavior is a strong motivation to perform series of heavy ion experiment under different circumstances.

\section{Experimental results from RHIC to LHC}
\label{sec:expresults}

In this section, I will concentrate on the $\alpha$ parameter from RHIC to LHC energies. However, the other L\'evy parameters are also important and interesting, they are out of the scope of the recent review. For detailed descriptions I would like to guide the interested reader to the original papers in Refs \cite{PHENIX:2017ino,Kincses:2019czd,Kincses:2018vuo,Lokos:2018dqq,Porfy:2019scf,PorfyNA61,CMS:2022cvh} and references in them. More detailed experimental review papers are also available in Refs. \cite{Csanad:2017usp,Csanad:2020xbf}.

\begin{figure}
    \centering
    \includegraphics[width=0.7\textwidth]{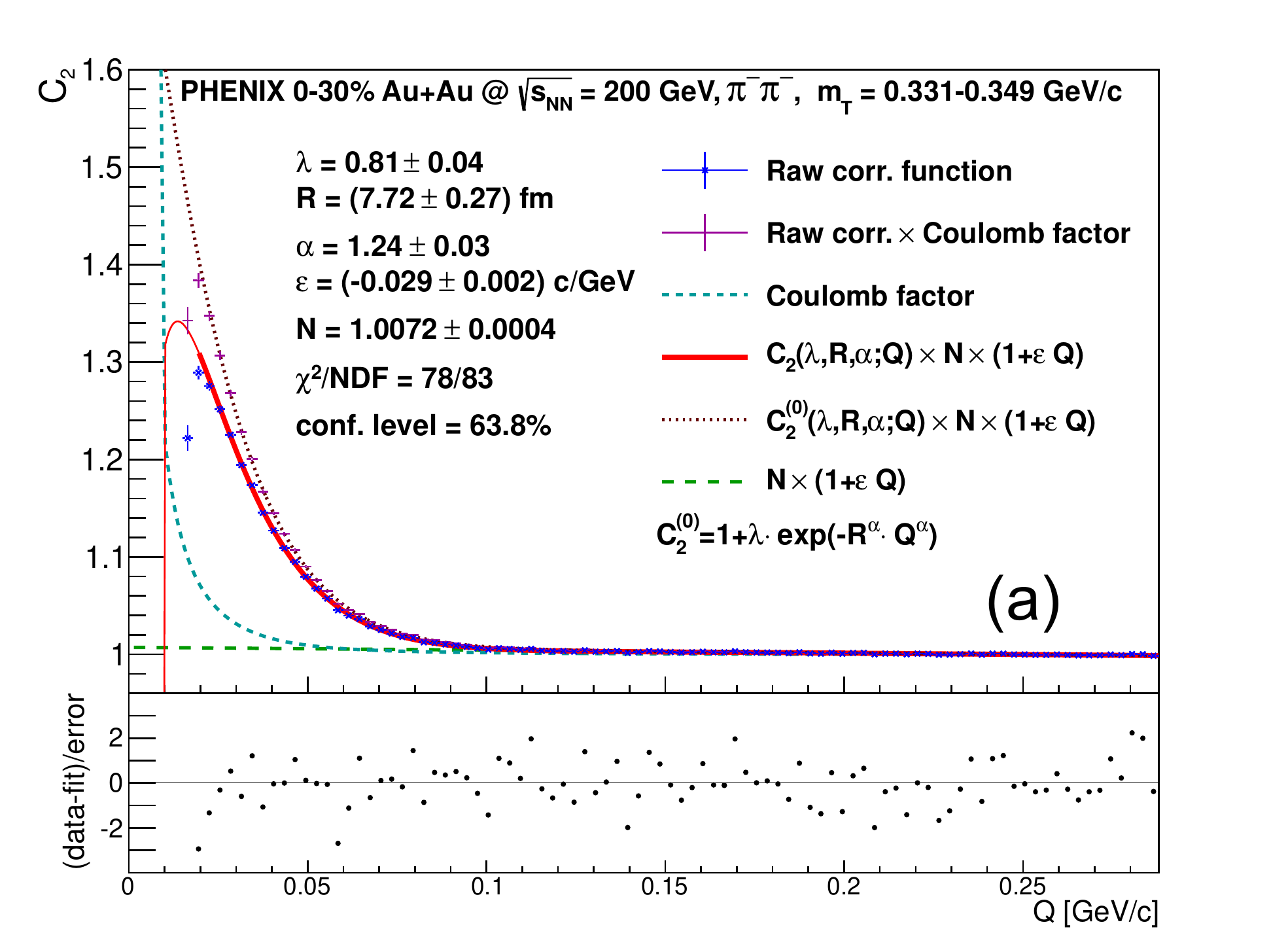}
    \caption{Example fit of a two particle correlation function and the values of the L\'evy parameters in PHENIX at $\sqrt{s_\textmd{NN}}= 200$ GeV.}
    \label{fig:exampleC2}
\end{figure}

The first indication of the deviance from the Gaussian shape was published by the PHENIX collaboration in Ref. \cite{PHENIX:2017ino}. The correlation function were measured in Au+Au collisions at $\sqrt{s_\textmd{NN}}=200$ GeV center-of-mass energy. An example fit of the two-particle correlation function can be seen in Fig. \ref{fig:exampleC2}. It could be emphasised that the fit is statistically acceptable, and the $\alpha$ parameter deviates significantly from the Cauchy and Gaussian case.

\begin{figure}
    \centering
    \includegraphics[width=0.49\textwidth]{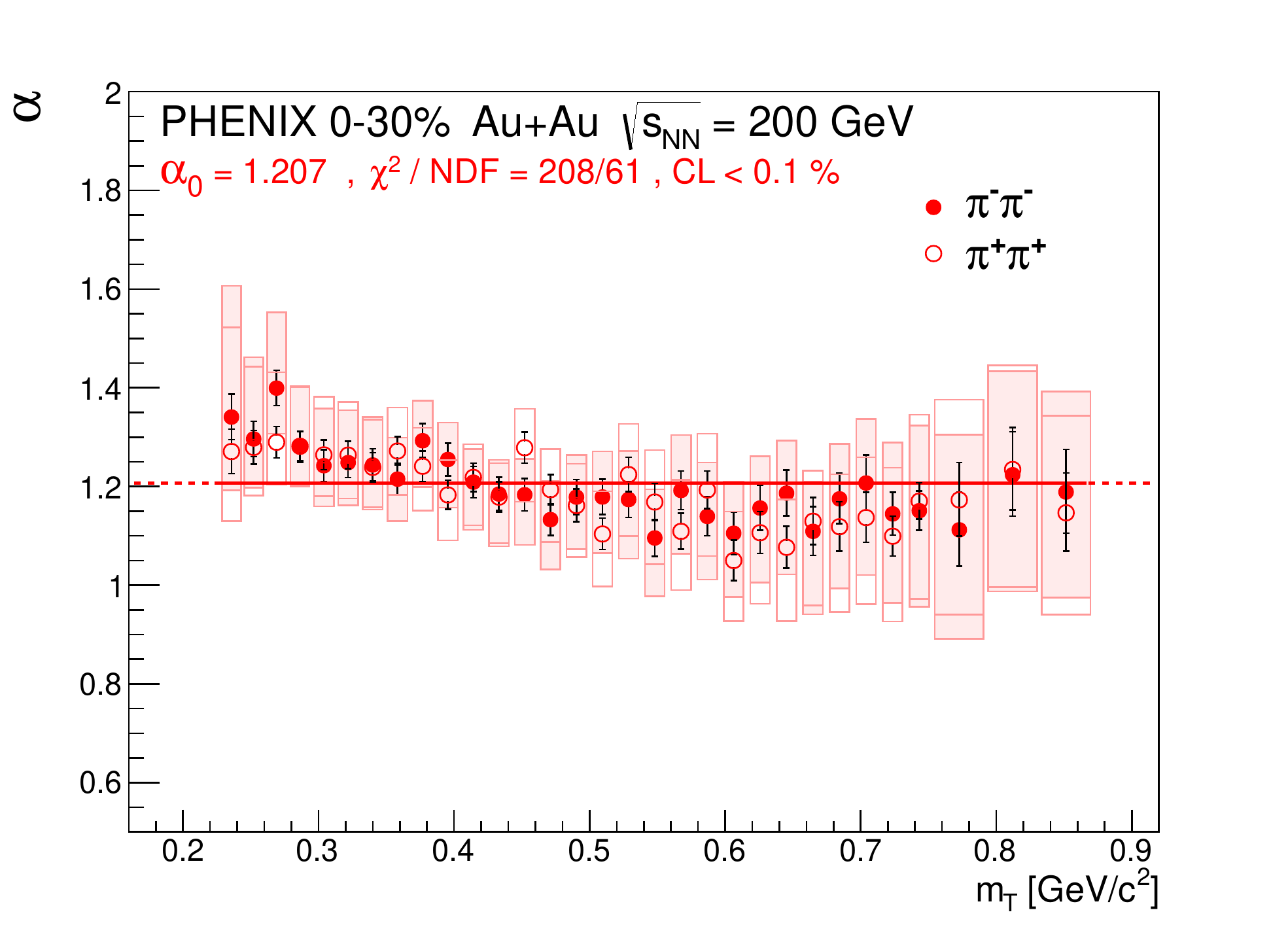}
    \includegraphics[width=0.49\textwidth]{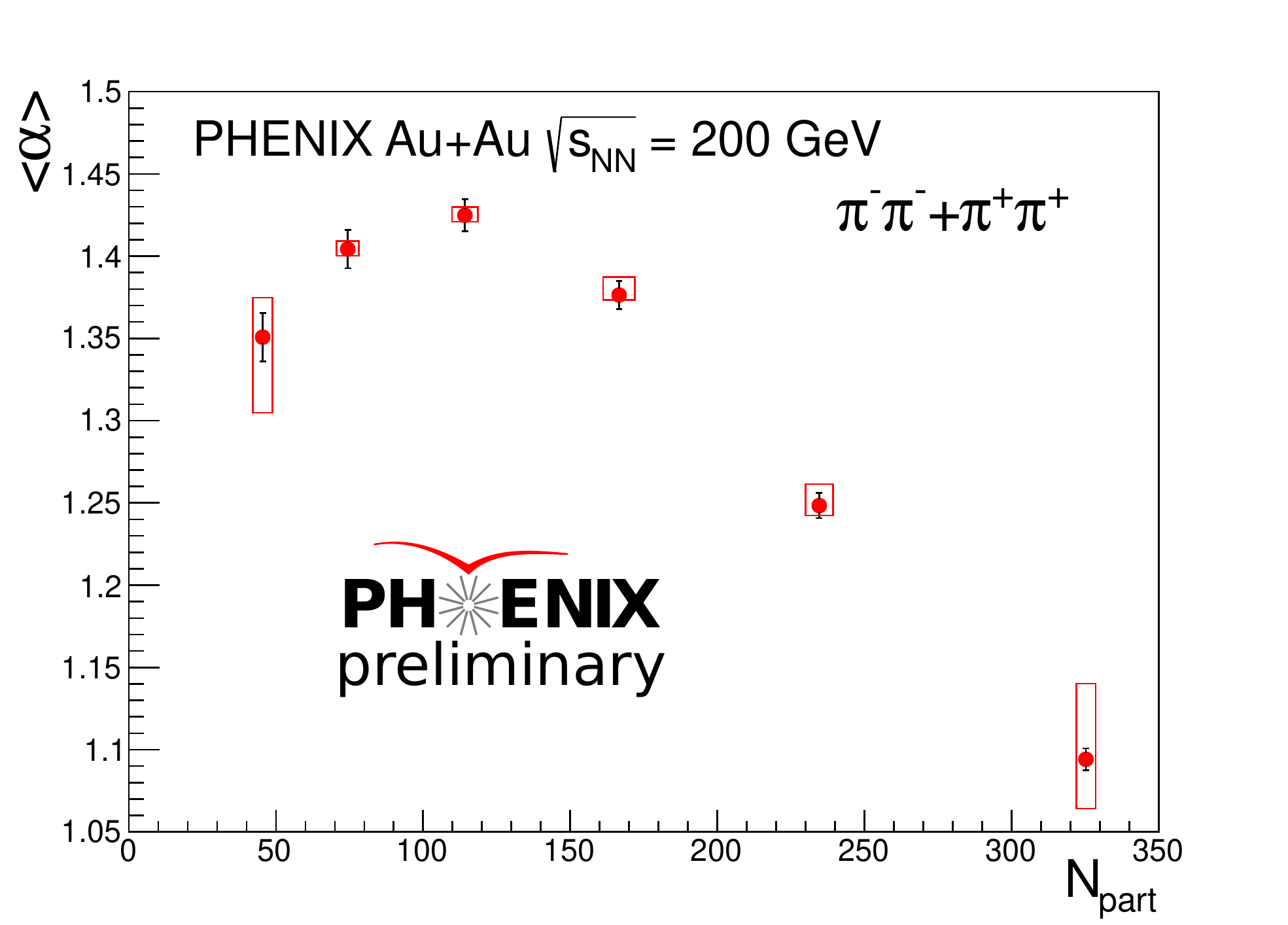}
    
    \includegraphics[width=0.49\textwidth]{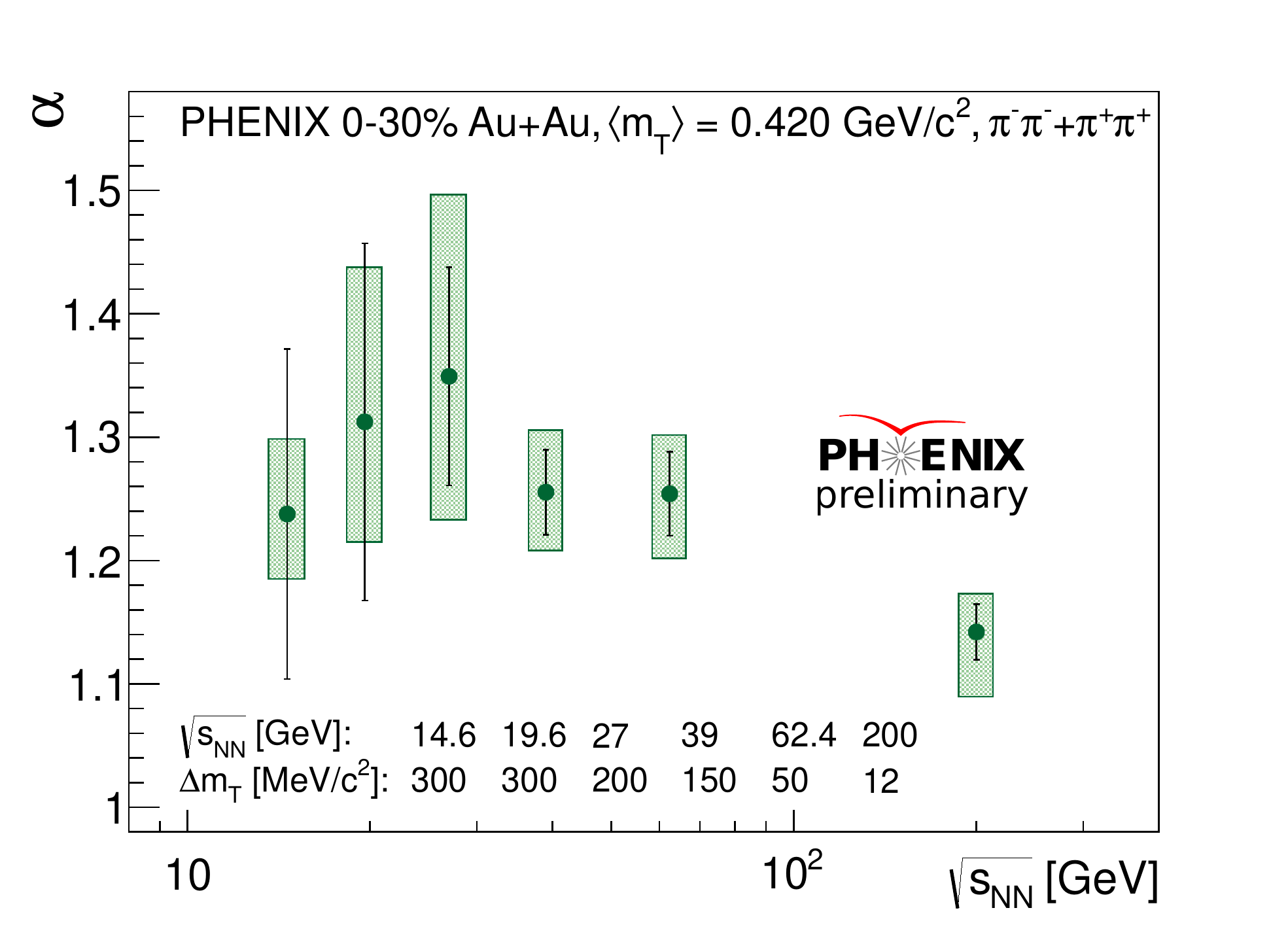}
    \caption{The transverse mass ($m_\textmd{T}$), the centrality ($N_\textmd{part}$) and the center-of-mass energy ($\sqrt{s_\textmd{NN}}$) dependence of the L\'evy $\alpha$ parameter at PHENIX in Au+Au collisions from 15 GeV to 200 GeV. Due to the lack of statistics at lower energies, only one transverse mass bin was considered at each energy. The width of these bins are shown in the figure. IT can be observed that in the investigated cases $1<\alpha<2$. }
    \label{fig:alpha_PHENIX_200GeV}
\end{figure}

The $\alpha$ parameter was measured at this energy as a function of the transverse mass $m_\textmd{T}$ of the pair with 0-30\% centrality selection, the $m_\textmd{T}$ averaged values for different, 10\% wide centrality ranges were determined. The center-of-mass energy dependence was also investigated. These can be seen in Fig. \ref{fig:alpha_PHENIX_200GeV}. The centrality dependent preliminary results can be found in Ref. \cite{Lokos:2018dqq}.

It can be seen from Fig. \ref{fig:alpha_PHENIX_200GeV}. that the L\'evy index indicates that the shape of the correlation function nor Cauchy neither Gaussian, moreover depends on the centrality. It is also far from the theoretically predicted critical value at top RHIC energies. The $\alpha$ parameter, however, does not exhibit strong dependence on $\sqrt{s_\textmd{NN}}$, as it is shown in Fig. \ref{fig:alpha_PHENIX_200GeV}. Due to the lack of statistics, only one transverse mass bin was considered at each energy. What is remarkable from the results is that the $\alpha$ is not consistent with either special values at any energies.

\begin{figure}
    \centering
    \includegraphics[width=0.9\textwidth]{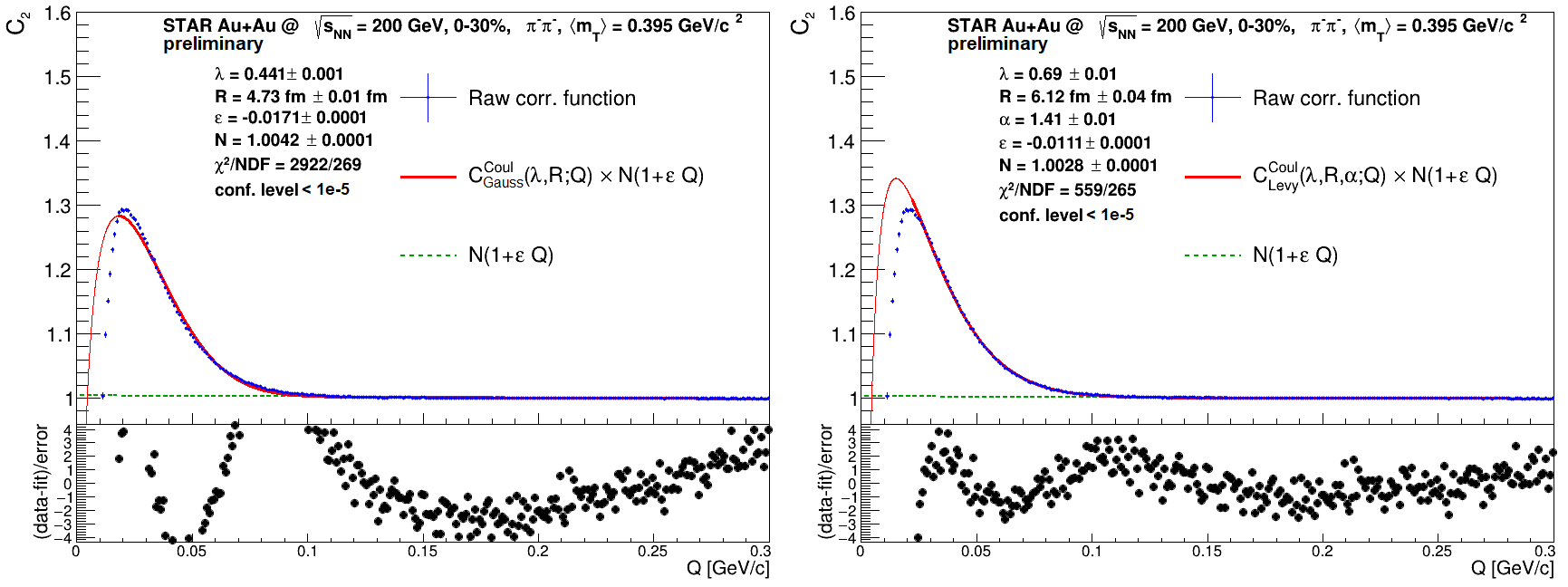}
    \caption{Example fits from RHIC STAR at $\sqrt{s_\textnormal{NN}}= 200$ GeV. The Gaussian fit (left) gives a worse description of the measured correlation function then the L\'evy fit (right), however, in terms of $\chi^2/\textmd{NDF}$ none of them are acceptable.}
    \label{fig:GaussLevy_STAR_200GeV}
\end{figure}

In the RHIC STAR experiment, the Bose-Einstein correlation functions can be measured with enhanced precision compared to PHENIX. The preliminary results showed that the shape of the correlation function cannot be characterized with Gaussian but the L\'evy assumption, although gives a better description in terms of $\chi^2/\textmd{NDF}$, cannot fully characterize the measured data \cite{Kincses:2019czd}. This can be seen in Fig. \ref{fig:GaussLevy_STAR_200GeV}.

\begin{figure}
    \centering
    \includegraphics[width=0.49\textwidth]{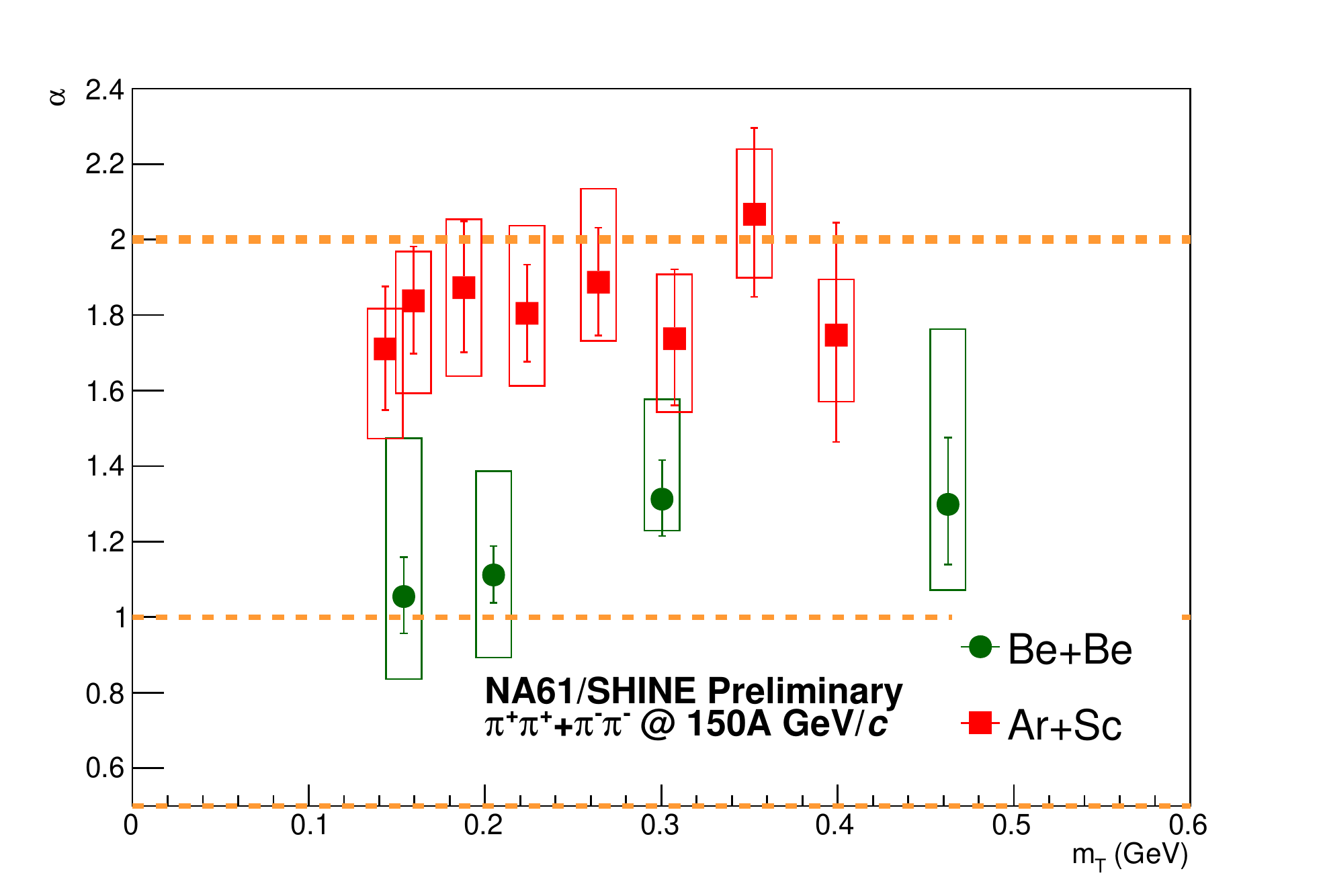}
    \includegraphics[width=0.49\textwidth]{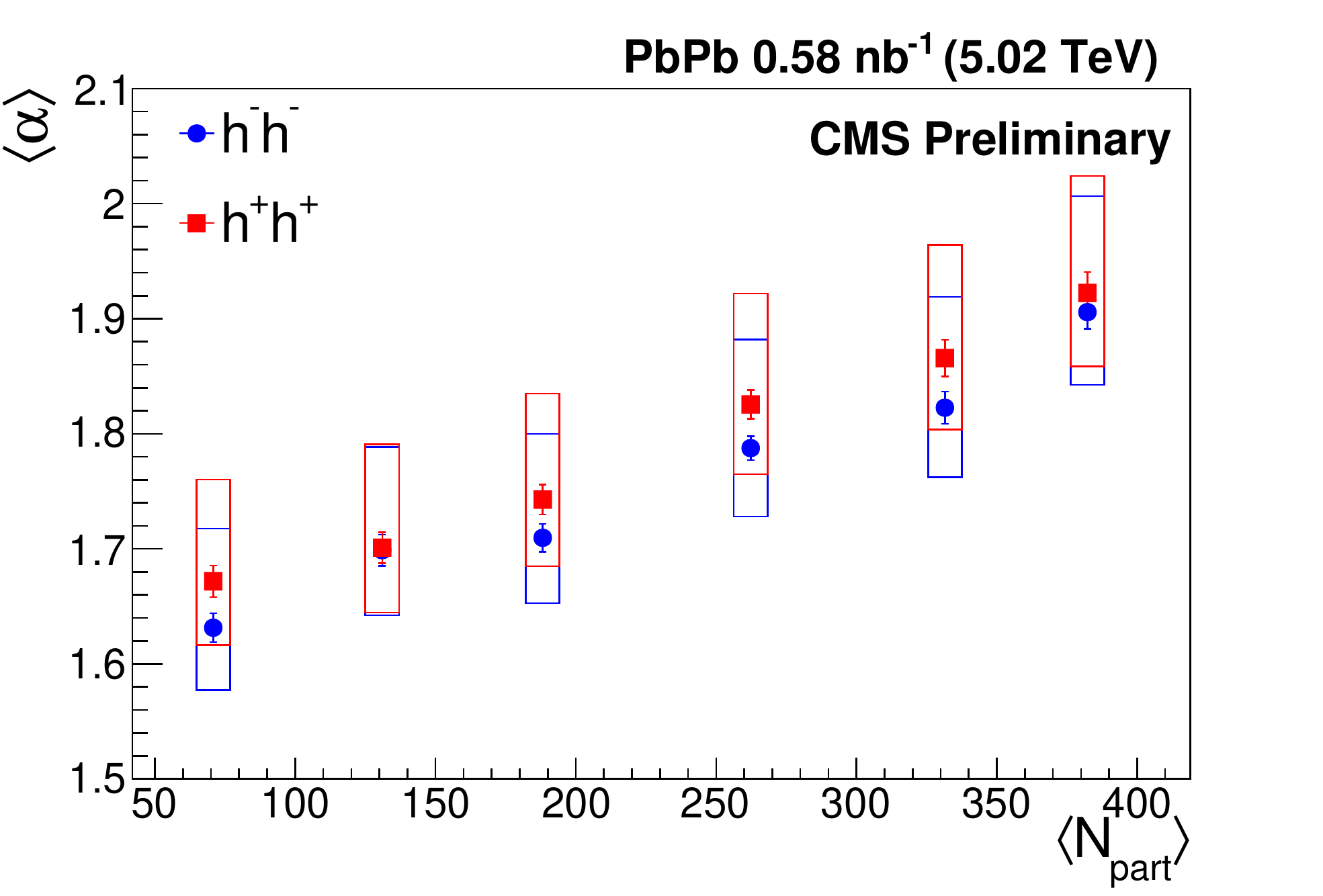}
    \caption{The $\alpha(m_\textmd{T})$ measured at NA61/SHINE experiment in Be+Be and in Ar+Sc collisions at $\sqrt{s_\textmd{NN}}= 150$ AGeV (left) and the CMS measurements  in Pb+Pb collision at $\sqrt{s_\textmd{NN}}= 5.02$ TeV (right).}
    \label{fig:NA61_CMS_alpha}
\end{figure}

Correlation measurements were performed at CERN lower energies and lighter systems with the NA61/SHINE experiment reported in Refs. \cite{Porfy:2019scf,PorfyNA61}. The Bose-Einstein measurements were performed in Be+Be and Ar+Sc systems at 150 AGeV energy per nucleons ($\sqrt{s_\textnormal{NN}} \sim 17$ GeV). Bose-Einstein correlation functions can be measured at high energy with the CMS experiment, as it was presented on the recent Quark Matter presentation (Ref. \cite{CMS:2022cvh}). In NA61/SHINE and in CMS $1<\alpha<2$ gives a good description of the measured Bose-Einstein correlation functions, the L\'evy assumption is the statistically acceptable choice as it is shown in Fig. \ref{fig:NA61_CMS_alpha} The system size dependence was already observed in PHENIX since centrality dependence can be translated into system size dependence, however, in the Ar+Sc collisions the $\alpha$ has slightly higher values then in RHIC but still below 2. This higher values were observed at the CMS experiments as well.

\section{Conclusions}
\label{sec:conculsion}

Bose-Einstein correlation measurements were performed in various colliding systems at various center-of-mass energies in RHIC and LHC experiments, and the L\'evy parametrization was used to characterize them. The system size and $m_\textmd{T}$ dependencies of the parameters were explored. It was observed that the shape of the correlation function cannot be characterized with Gaussian nor with Cauchy shape, instead the general L\'evy form should be utilized. The experimental observations suggest that this value is in the range of $1 < \alpha < 2$ at the investigated energies and systems, i.e., no indication for critical behavior was observed so far but the precise shape analyses of the correlation functions would help to understand other physical processes besides criticality.

\bibliographystyle{prlsty}
\bibliography{main.bib}

\begin{thebibliography}{10}

\bibitem{HanburyBrown:1956bqd}
R. Hanbury~Brown and R.~Q. Twiss, Nature {\bf 178},  1046  (1956).

\bibitem{Lednicky:2001qv}
R. Lednick{\'y},  [arXiv:nucl-th/0112011].

\bibitem{Kurgyis:2019xzt}
B. Kurgyis, Phys. Part. Nucl. {\bf 51},  263  (2020) [arXiv:1910.05019].

\bibitem{Yano:1978gk}
F.~B. Yano and S.~E. Koonin, Phys. Lett. {\bf 78B},  556  (1978).

\bibitem{Csorgo:1994in}
T. Cs{\" o}rg{\H o}, B. L{\"o}rstad, and J. Zim{\'a}nyi, Z. Phys. C {\bf 71},
  491  (1996) [arXiv:hep-ph/9411307].

\bibitem{1991PhLB27069B}
M.~G. {Bowler}, Physics Letters B {\bf 270},  69  (1991).

\bibitem{Akkelin:1995gh}
S.~V. Akkelin and {\relax Yu}.~M. Sinyukov, Phys. Lett. {\bf B356},  525
  (1995).

\bibitem{PHENIX:2017ino}
A. Adare {\it et~al.}, Phys. Rev. C {\bf 97},  064911  (2018)
  [arXiv:1709.05649].

\bibitem{Csanad:2019lkp}
{Csan\'ad, M. and L{\"o}k{\"o}s, S. and Nagy, M.}, Phys. Part. Nucl. {\bf 51},
  238  (2020) [arXiv:1910.02231].

\bibitem{Kincses:2019rug}
D. Kincses, M.~I. Nagy, and M. Csan\'ad, Phys. Rev. C {\bf 102},  064912
  (2020) [arXiv:1912.01381].

\bibitem{Csanad:2020qtu}
M. Csan{\'a}d {\it et~al.},  [arXiv:2007.07167].

\bibitem{Kincses:2022eqq}
D. Kincses, M. Stefaniak, and M. Csan\'ad, Entropy {\bf 24},  308  (2022)
  [arXiv:2201.07962].

\bibitem{Csorgo:2005it}
T. Cs{\"o}rg{\H o}, S. Hegyi, T. Nov{\'a}k, and W.~A. Zajc, AIP Conf. Proc.
  {\bf 828},  525  (2006) [arXiv:nucl-th/0512060].

\bibitem{Stephanov:1998dy}
M.~A. Stephanov, K. Rajagopal, and E.~V. Shuryak, Phys. Rev. Lett. {\bf 81},
  4816  (1998) [arXiv:hep-ph/9806219].

\bibitem{PhysRevB.52.6659}
H. Rieger, Phys. Rev. B {\bf 52},  6659  (1995).

\bibitem{Kincses:2019czd}
D. Kincses, Phys. Part. Nucl. {\bf 51},  267  (2020) [arXiv:1911.05352].

\bibitem{Kincses:2018vuo}
D. Kincses, Acta Phys. Polon. Supp. {\bf 12},  445  (2019) [arXiv:1811.08311].

\bibitem{Lokos:2018dqq}
S. L{\"o}k{\"o}s, Universe {\bf 4},  31  (2018) [arXiv:1801.08827].

\bibitem{Porfy:2019scf}
B. P\'orfy, Universe {\bf 5},  154  (2019) [arXiv:1906.06065].

\bibitem{PorfyNA61}
B. P{\'o}rfy, Quark Matter poster,
  https://indico.cern.ch/event/895086/contributions/4703951/.

\bibitem{CMS:2022cvh}
B. K{\'o}rodi, CMS-PAS-HIN-21-011 (2022).

\bibitem{Csanad:2017usp}
M. Csan{\'a}d, Universe {\bf 3},  85  (2017) [arXiv:1711.05575].

\bibitem{Csanad:2020xbf}
M. Csan{\'a}d, J. Phys. Conf. Ser. {\bf 1602},  012009  (2020)
  [arXiv:2007.04751].

\end{thebibliography}

\end{document}